\documentclass[preprint,12pt]{elsarticle}
\biboptions{square} 

 \usepackage{graphicx}

\usepackage{amssymb}
 \usepackage{amsthm}
 \usepackage{amsmath}
 \usepackage{caption}
 \usepackage{subcaption}
 \usepackage{float}
 \usepackage{multirow}
 \usepackage{epstopdf}
 \usepackage{booktabs}
 \usepackage{float}
 
\usepackage{color}
\usepackage{ulem}

\newcommand{\CRe}[1]{}

\usepackage[labelformat=simple]{subcaption}
\graphicspath{{./plots/}}

\newcommand{\IMF}{\mathrm{IMF}}

\biboptions{comma}

\journal{}

\begin{document}

\begin{frontmatter}



\title{
Dynamic correlations at different time-scales with Empirical Mode Decomposition
}


\author[a,b]{Noemi Nava}
\author[a,b,c,d]{T. Di Matteo}
\author[a,b]{Tomaso Aste}
\ead{t.aste@ucl.ac.uk}
\address[a]{Department of Computer Science, University College London, \\ Gower Street, London, WC1E 6BT, UK}
\address[b]{Systemic Risk Centre, \\ London School of Economics and Political Sciences, \\ London, WC2A2AE, UK}
\address[c]{Department of Mathematics, King's College London, \\ The Strand, London, WC2R 2LS, UK}
\address[d]{Complexity Science Hub \\ Josefstaedter Strasse 39, A 1080 Vienna, Austria.
}

\begin{abstract}
The Empirical Mode Decomposition (EMD) provides a tool to characterize time series in terms of its implicit components  oscillating at different time-scales. 
We apply this decomposition to intraday  time series of the following three financial indices: the S\&P 500  (USA),  the IPC (Mexico) and the VIX (volatility index USA), obtaining time-varying multidimensional cross-correlations at different time-scales. 
The correlations computed over a rolling window are compared across the three indices, across the components at different time-scales, at different lags and over time. 
We uncover a rich heterogeneity of interactions which depends on the time-scale and has important led-lag relations which can have practical use for portfolio management, risk estimation and investments.
\end{abstract}


\end{frontmatter}

\section{Correlation structures in financial time series}
Financial   time series are correlated and the structure of these correlations reflects important market properties \cite{mantegna1999introduction,aste2010correlation,aste2006dynamical,aste2010introduction}. 
Financial markets operate at different time horizons \cite{di2007multi},  and  characterizing the relation between  market prices at different time-scales is essential to capture the complexity of market dynamics for portfolio management, risk management and investments \cite{buonocore2017asymptotic,buonocore2016two,musmeci2014risk,musmeci2016interplay}.
It is well established and documented that  correlations between stock returns varie over time (see for instance \cite{aste2010correlation,Longin19953,Rua2009632,Vacha2012241}).
It is instead less understood and established how correlations between financial assets  vary over time-scales \cite{tumminello2007correlation}. 
Most studies  only focus on a specific  time-scale. 
However, changes of correlation at different time-scales  have important practical consequences. Indeed, if the correlation between two assets varies across time-scales, then market participants with short and long term-horizons have different risk exposures and must adapt their strategies according to the different parts of the correlation spectrum. 
Furthermore, investigating  both  the time dependent and the time-scale dependent dynamics of correlations can provide insights on the collective behaviour of traders with varying strategies \cite{Longin19953,bartolozzi2007multi}.
This is the topic of the present paper where we use a simple methodology  to perform this research.

Our approach  is similar to what was  recently introduced  by Chen et al. \cite{Chen2010233} who proposed to use the Empirical Mode Decomposition (EMD) to estimate the so-called Time-Dependent  Intrinsic Correlation (TDIC). 
In this approach, two time series are first decomposed into a set of  components called  Intrinsic Mode Functions (IMFs)  oscillating at different time scales. Then, the Pearson correlation is calculated in  an adaptive window whose length depends on the instantaneous period of the   IMFs.
In this paper, we introduce a simplified version of this approach applied to intra-day data (30 seconds) of three indices: the S\&P 500  (USA), the IPC (Mexico) and the VIX (volatility index USA). We  compute cross-correlations and lagged cross-correlations from different IMFs generated from rolling windows.
This yields to dynamic cross-correlations across time scales.
The results uncover the presence of  cross-scale coupling between the time series and identify some relevant led-lag relation at specific time-scales which could be relevant for practical purposes in portfolio management. 

Another technique to measure time-varying correlation and  which provides similar outputs is the wavelet coherence \cite{Rua2009632, Vacha2012241}.
However, differently from the wavelet transform,  the EMD does not require any a priori filter function \cite{Peng2005}. The EMD relies on less assumptions, it is a fully data-driven decomposition which can be applied to non-stationary and non-linear data \cite{Huang}. 
A similar framework was proposed in \cite{horvatic2011detrended} where  correlations between time series with periodic trends were estimated by using local piecewise polynomial detrending.

This paper is organized as follows. 
In Section \ref{sec:EMD},  we introduce the basic concepts of the EMD and the IMF. 
In Section \ref{sec:GlobalPearson}, the computation of  cross-correlations across time-scales, time-lags and time-windows is described. 
 Section \ref{sec:CorrHFD} reports the application to real data on   three indices: the S\&P 500  (USA), the IPC (Mexico) and the VIX.
 The discussions and conclusions are provided in Section \ref{sec:CorrConclusion}.

\section{Empirical mode decomposition (EMD)}
\label{sec:EMD}

The EMD method identifies a finite set of  oscillations with scale defined by the local maxima and the local minima of the data itself.	Each oscillation is	 empirically derived from the data and	is referred to as an Intrinsic Mode Function (IMF).
An IMF must satisfy two criteria \cite{Huang}:
\begin{enumerate}
\item The number of extrema and the number of zero crossings must either be equal or differ at most by one.
\item At any point, the mean value of the envelope defined by the local maxima and the envelope defined by the local minima is zero. 
\end{enumerate}

The first condition forces an IMF to be a narrow-band signal with  no riding waves. The second condition ensures that the instantaneous time-scale will not have fluctuations arising from an asymmetric wave form \cite{Huang}.	
The IMFs are obtained through a	 process called sifting process	 which	uses  local extrema to separate oscillations starting with the highest time-scale. Given a time series $X(t)$, $t=1,2,...,T$, the process decomposes it into a finite number of components, called Intrinsic Mode Functions, here denoted as $\IMF_k(t)$, $k= 1,..., n$, and a residue $r_n(t)$.	The residue is	the non-oscillating drift of the data.
If the decomposed data consists of uniform scales in the time-scale space, the EMD acts as a dyadic filter and the total number of IMFs is approximately equal to n$=\log_2(T)$ \cite{Flandrin}.	
 At the end of the decomposition  process, the original time series can be reconstructed as:

\begin{equation}
X(t)=\sum_{k=1}^n  \IMF _k(t) + r_n(t).
\label{eq:EMDe}
\end{equation}	

\begin{figure}[h!]
	  \centering
	  \begin{subfigure}[b]{0.49\textwidth}
			  \includegraphics[width=\textwidth]{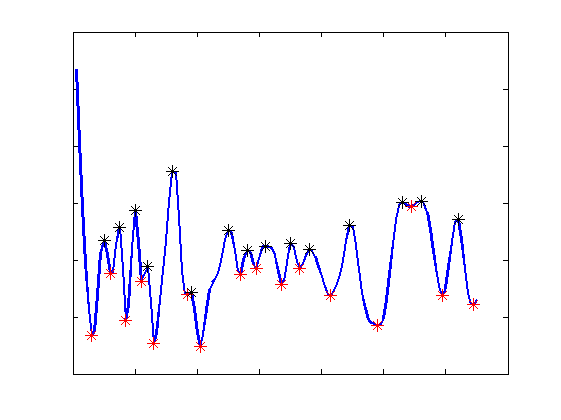}
			  \caption{Local maxima and minima.}
			  \label{fig:siftingA}
	  \end{subfigure}%
	  \begin{subfigure}[b]{0.49\textwidth}
			  \includegraphics[width=\textwidth]{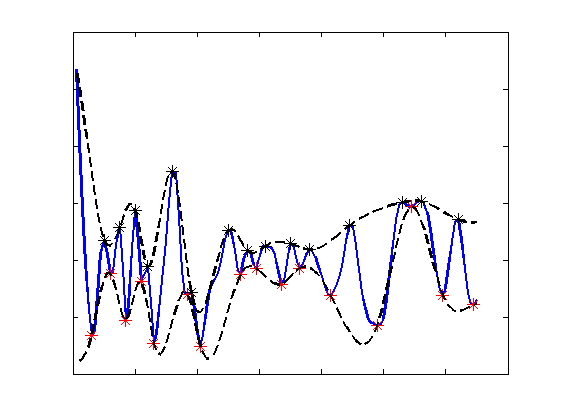}
			  \caption{Upper and lower envelopes.}
			  \label{fig:siftingB}
	  \end{subfigure}  
		  \newline
	  \begin{subfigure}[b]{0.49\textwidth}
			  \includegraphics[width=\textwidth]{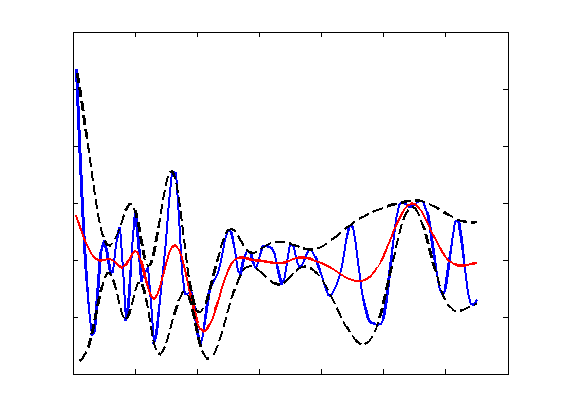}
			  \caption{Envelope mean.}
			  \label{fig:siftingC}
	  \end{subfigure}
		\begin{subfigure}[b]{0.49\textwidth}
					  \includegraphics[width=\textwidth]{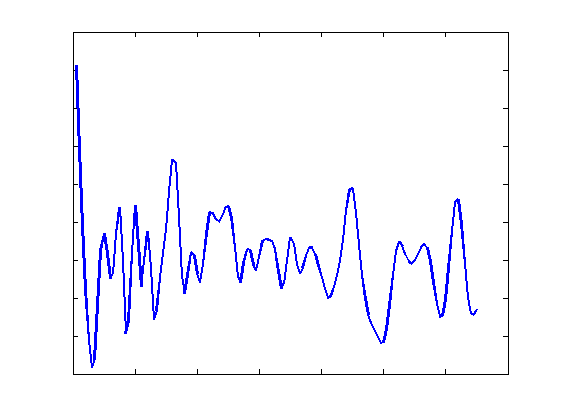}
					  \caption{Time series after one sifting step.}
					  \label{fig:siftingD}
			  \end{subfigure}
	  \caption{Example of one sifting step in the construction of a IMF. 
	  (a)	Input time series highlighting the local maxima and the local minima.
	  (b)	Time series with the interpolated upper and lower envelopes (slashed lines).
	  (c)	Time series with the envelopes and the mean of both envelopes (red line).
	  (d)	First iteration of the sifting process. 
	  In this example, the extracted function does not satisfy the IMF's conditions and therefore another set of sifting processes must be applied (see Fig.\ref{fig:sifting2}).
	  }
 \label{fig:sifting}
\end{figure}

\begin{figure}[h!]
	 \centering
	 \begin{subfigure}[b]{0.49\textwidth}
			 \includegraphics[width=\textwidth]{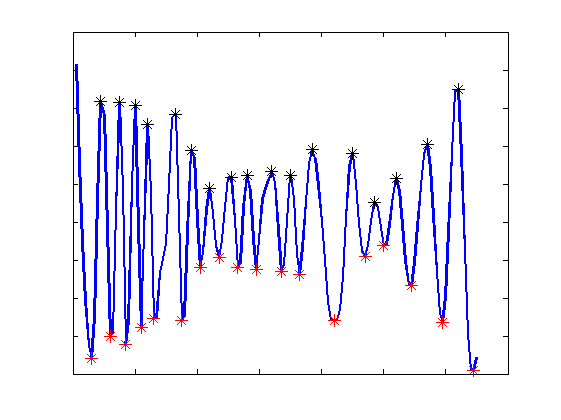}
			 \caption{Local maxima and minima.}
			 \label{fig:sifting2A}
	 \end{subfigure}%
	 \begin{subfigure}[b]{0.49\textwidth}
			 \includegraphics[width=\textwidth]{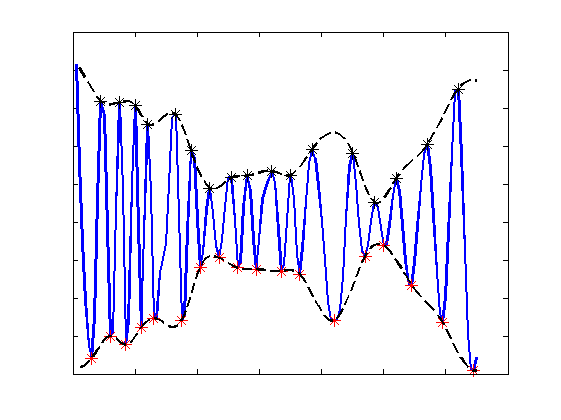}
			 \caption{Upper and lower envelopes.}
			 \label{fig:sifting2B}
	 \end{subfigure}  
		 \newline
	 \begin{subfigure}[b]{0.49\textwidth}
			 \includegraphics[width=\textwidth]{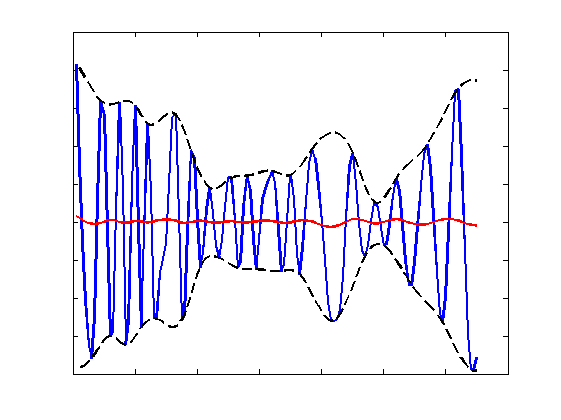}
			 \caption{Envelope mean.}
			 \label{fig:sifting2C}
	 \end{subfigure}
	   \begin{subfigure}[b]{0.49\textwidth}
					 \includegraphics[width=\textwidth]{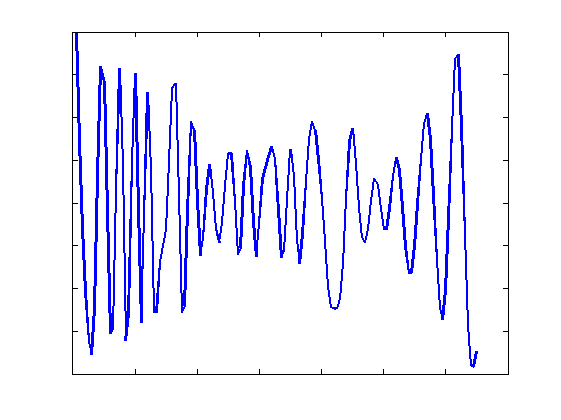}
					 \caption{IMF example.}
					 \label{fig:sifting2D}
			 \end{subfigure}
	 \caption{\label{fig:sifting2} Example of the few final sifting steps which produce a valid IMF.  
	 (a)	Input time series highlighting the local maxima and the	 local minima.
	 (b)	Input time series with the interpolated upper and lower envelopes (slash lines).
	 (c)	Input time series with the envelopes and  the mean of both envelopes (red line).
	 (d)	Last iteration of the sifting process, the extracted function is the first IMF.
	 }
\end{figure}

The	 EMD is implemented through the following steps \cite{Huang}:
\begin{enumerate}
\item Initialize the residue to the original time series $r_0(t)=X(t)$ and set the IMF index $k=1$.

\item Extract the $k^\textrm{th}$ IMF:

\begin{enumerate}
\item initialize $h_0(t)= r_{k-1}(t)$ and set the iteration counter $i=1$;

\item find the local maxima and the local minima  of $h_{i-1}(t)$ (see Figure \ref{fig:siftingA}); 

\item create the upper envelope $E_u(t)$ by interpolating between the local maxima and, analogously, create lower envelope $E_l(t)$ by interpolating the local minima (see Figure \ref{fig:siftingB});  

\item calculate the mean of both envelopes as $m_{i-1}(t) = \frac{ E_u(t) + E_l(t)}{2}$ (see   Figure \ref{fig:siftingC});

\item subtract the mean envelope  from the input time series, obtaining	$h_{i}(t)= h_{i-1}(t)- m_{i-1}(t)$, see Figure \ref{fig:siftingD}; 

\item verify if $h_{i}(t)$	satisfies the IMF's conditions:

\begin{itemize}
\item if $h_{i}(t)$ does not satisfy the IMF's conditions, increase $i=i+1$ and repeat the sifting process from step (b) (see Figure \ref{fig:siftingD}); 

\item if $h_{i}(t)$ satisfies the IMF's conditions, set	 $\IMF_k(t)= h_{i}$ and define	$r_k(t)= r_{k-1}(t)-\IMF_k(t)$  (see	 Figure \ref{fig:sifting2}).

\end{itemize}
\end{enumerate}

\item When the residue	$r_k(t)$ is either a constant, a monotonic slope or contains only one extrema  stop the	 process, otherwise continue the decomposition from step $2$, setting $k=k+1$.

\end{enumerate}

In Figure \ref{fig:sifting}, we exemplify some steps of the sifting process. 
After one iteration of the sifting process, the function $h_{1}(t)$  is obtained (Figure \ref{fig:siftingD}). 
In this example, the resulting function is not symmetric and does not have zero mean, hence it is not an IMF yet. 
Consequently, more iterations of the sifting processes need to be applied to extract the first IMF of the input time series.  
These further iterations are shown in Figures \ref{fig:sifting2A}, \ref{fig:sifting2B} and \ref{fig:sifting2C} with the last sifting iteration which extracts the first IMF, shown in Figure \ref{fig:sifting2D}. 

It must be noted that the EMD is based on the timescale separation and does not impose orthogonality, implying that in general the  sum of the variance of the components and the residue differs from  the variance of the input time series. However, in most practical cases, the difference is small \cite{Huang}. 

The sifting process eliminates the riding waves and smooths uneven amplitudes  \cite{Huang}. This process terminates when the local mean of the extracted IMF is zero. The difficulty is that this condition can only be approximated and in order to avoid over-sifting and converting	meaningful IMFs into meaningless fluctuations with constant amplitude, a stopping criterion needs to be implemented. 

\section{Cross-correlations on IMF}
\label{sec:GlobalPearson}

Let us consider two time series  $X(t)$ and $Y(t)$ with $t=1,2,	 \ldots, T$, with equal length $T$  and with equal intervals of time $s$ between observations. 

\subsection{Cross-correlations across time-scales}
The proposed time-scale-dependent correlation computes  the Pearson correlation coefficients between two components $\IMF_i^X$, $\IMF_j^Y$, $i,j=1,\ldots,n$ obtained from the decomposition of the time series $X(t)$ and $Y(t)$, respectively:
\begin{equation}
\rho^{XY}_{i,j}= \frac{1}{T}\sum\limits_{t=1}^{T}\frac{\left(\IMF_i^X(t) -\overline{\IMF_i^X}\right) \left(\IMF_j^Y(t)-\overline{\IMF_j^Y}\right)}{\sigma_i^X \sigma_j^Y}
\label{eq:IMFCORR}
\end{equation}
where  $\overline{\IMF_i^X} $ denotes the sample mean over time of $\IMF_i^X$ and  $\sigma_i^X$	  denotes the sample standard deviation of $\IMF_i^X$.

Although the IMFs are not theoretically stationary, the IMFs satisfy the  condition of having local mean equal to zero and can then be considered to be at least locally stationary \cite{Huang}. 
Contrary, the residue	 does not need to satisfy the IMF conditions, and particularly, for an initial non-stationary time series, the extracted residue will contain the trend of the time series, making it a non-stationary component. Thus, a correlation coefficient between residues	is just a measure of linear dependency of the  trends indicating if they move  in the same direction.  This correlation coefficient is likely to be high, and could give misleading results for the interpretation of the dependence structure.

\subsection{Time-dependent lagged cross-correlation at the same time-scale over a rolling window}
\label{sec:WlaggedCC}

We also compute lagged cross-correlations over  a rolling window, which for simplicity, we limit to the same time scale.
The cross-correlations between two different time series $X(t)$ and $Y(t)$,  lagged by $\lambda$,  over rolling windows of size $W$ and  at the same time-scale component $i$ is defined as: 
\begin{equation}
\rho^{XY}_{i}(t,\lambda) = \frac{1}{W-\lambda}\sum\limits_{\tau=t-W+1}^{t-\lambda}\frac{\left(\IMF_i^X(\tau)-\overline{\IMF_i^X}\right)\left(\IMF_i^Y(\tau+\lambda)-\overline{\IMF_i^Y}\right)}{\sigma_i^X \sigma_i^Y} \;\;.
\label{eq:WLCross}
\end{equation}
%
The time-lag $\lambda$ is measured in units of the sampling time-scale. %
The	 window	 approach has the advantage of only assuming local stationarity rather than stationarity over the entire time series.
Although this method is based on a simple measure of correlation (Pearson correlation), it adapts to the nature of the data and	 provides a dynamic measure of correlation across time-scales.

\section{Correlation analysis of  intraday financial data}
\label{sec:CorrHFD}
We consider intraday data sampled at 30-seconds intervals for	 two stock market indices and a volatility index, namely, the S\&P 500 index (USA),	 the IPC index (Mexico) and	   the VIX index (implied volatility index, calculated by the Chicago Board Options Exchange, USA). The  data was obtained from Bloomberg \cite{BloombergWeb}.
The observation period includes 184 days, ranging from September 2013 to July 2014 and it only considers the trading days	available for all the three indices. Each day has 780 data points (6.5 hours). 

\begin{figure}[h!]
	\centering	
\includegraphics[width=4.5 in, height= 2.3 in]{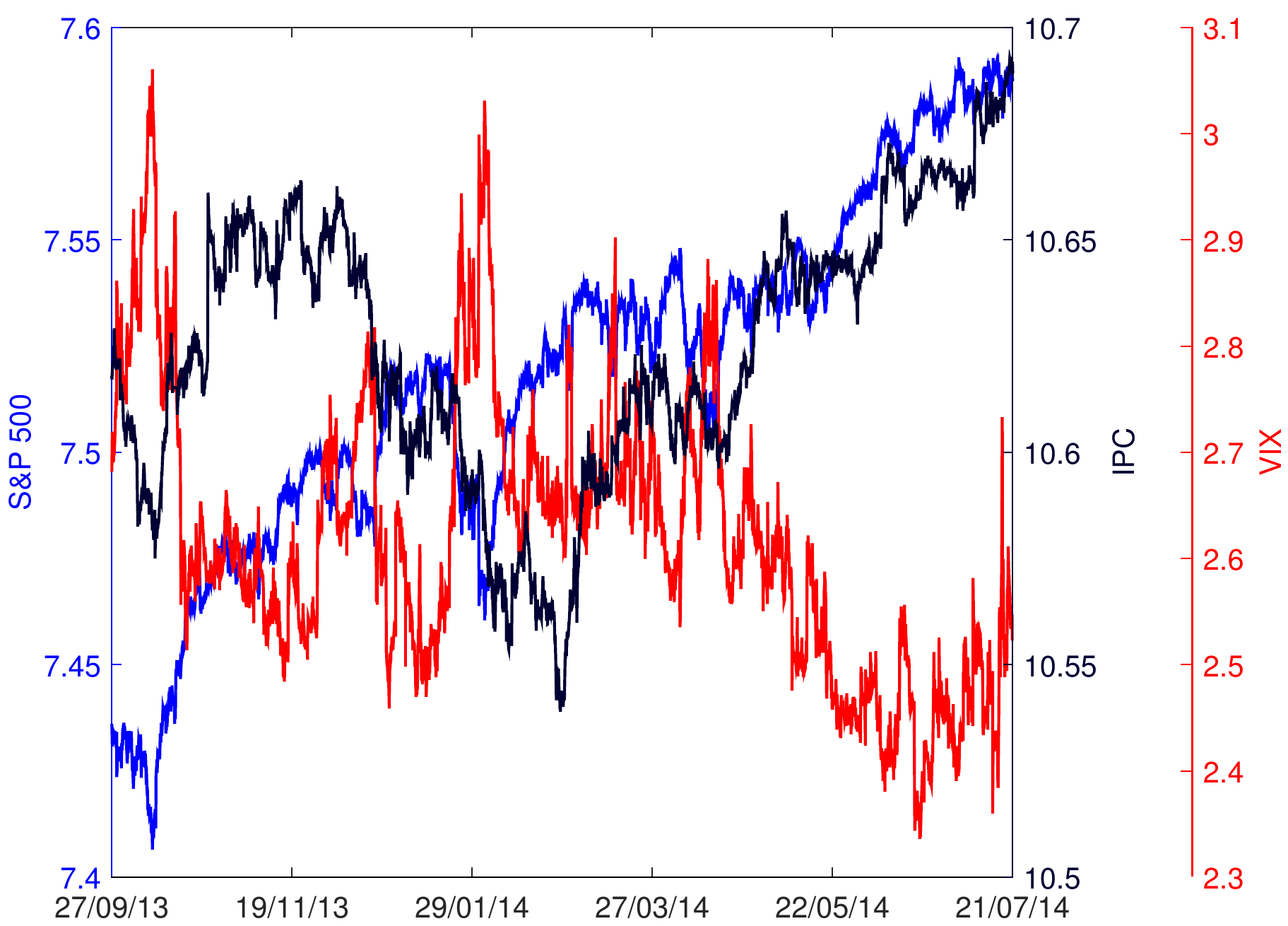}
			\caption{Intraday observations	(sampled at 30-seconds intervals) for the S\&P 500, the IPC and the VIX indices for the time period September 2013 to July 2014.}
			\label{fig:3TSLong}
 \end{figure}
Figure \ref{fig:3TSLong} reports the dynamics of these three indices over that time period. 
We can observe that     the S\&P 500 and the IPC indices have similar behaviours.
They are indeed positively correlated {with correlation coefficient between log-returns equal to 0.21}; this is in agreement with  previous studies \cite{Araujo,Mohamed}.	
On the contrary, the risk-price relationship between  the S\&P 500   and	the VIX indices shows negative correlation, as reported for example in  \cite{Whaley}.
  The correlation coefficient between these log-returns is equal to -0.26. 
Finally, the IPC  and the VIX indices are essentially uncorrelated with very small negative correlations (the correlation coefficient between log-returns equals to -0.02).

\subsection{Intraday analysis of correlation, example for the day July $18^{th}$ 2014}
\begin{figure}[h!]
	\centering	
\includegraphics[width=4.5 in, height= 2.3 in]{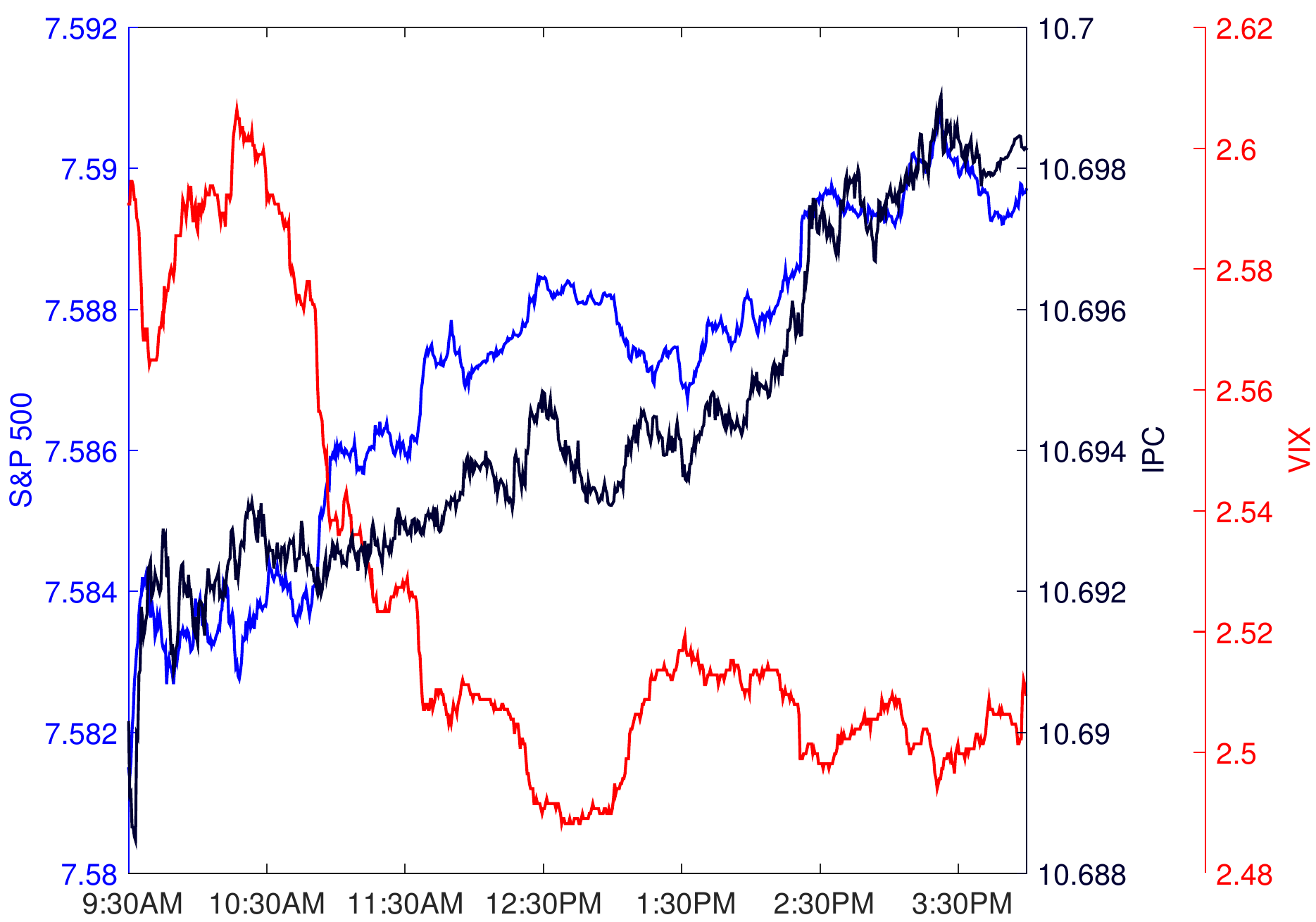}
			\caption{Intraday log-prices for the S\&P 500, the IPC and the VIX indices, example for July $18^{th}$ 2014.}
			\label{fig:3TS}
 \end{figure}

\begin{figure}[h!]
			  \centering
			  \begin{subfigure}[t]{0.33\textwidth}
						  \centering
					  \includegraphics[width=\textwidth, height= 2.6 in]{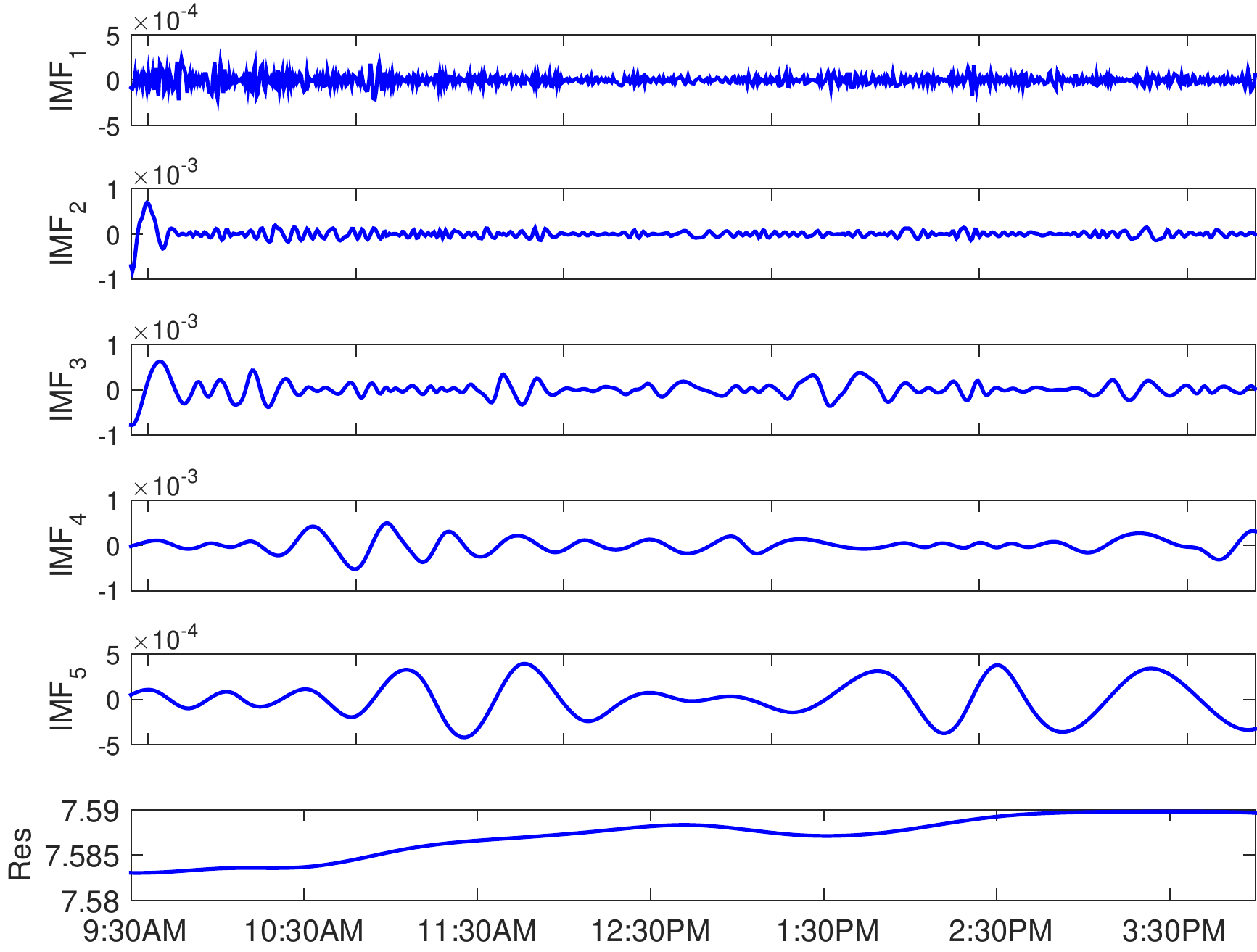}
					  \caption{S\&P 500 index.}
					  \label{fig:IMF1}
			  \end{subfigure}%
			  \begin{subfigure}[t]{0.33\textwidth}
						  \centering
					  \includegraphics[width=\textwidth, height= 2.6 in ]{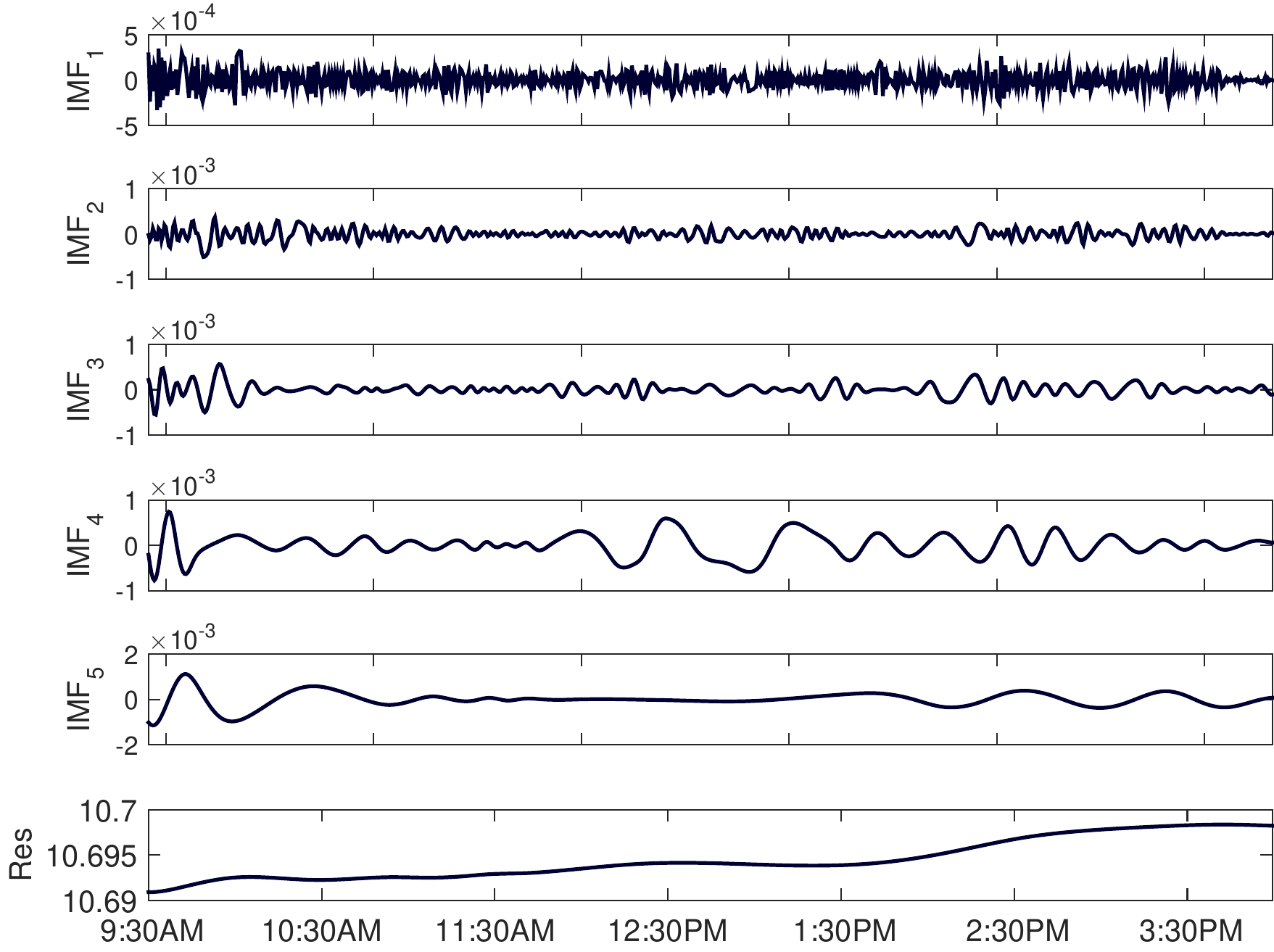}
					  \caption{IPC index.}
					  \label{fig:IMF2}
			  \end{subfigure} 
	 \begin{subfigure}[t]{0.33\textwidth}
						  \centering
						  \captionsetup{justification=centering}
					  \includegraphics[width=\textwidth, height= 2.6 in]{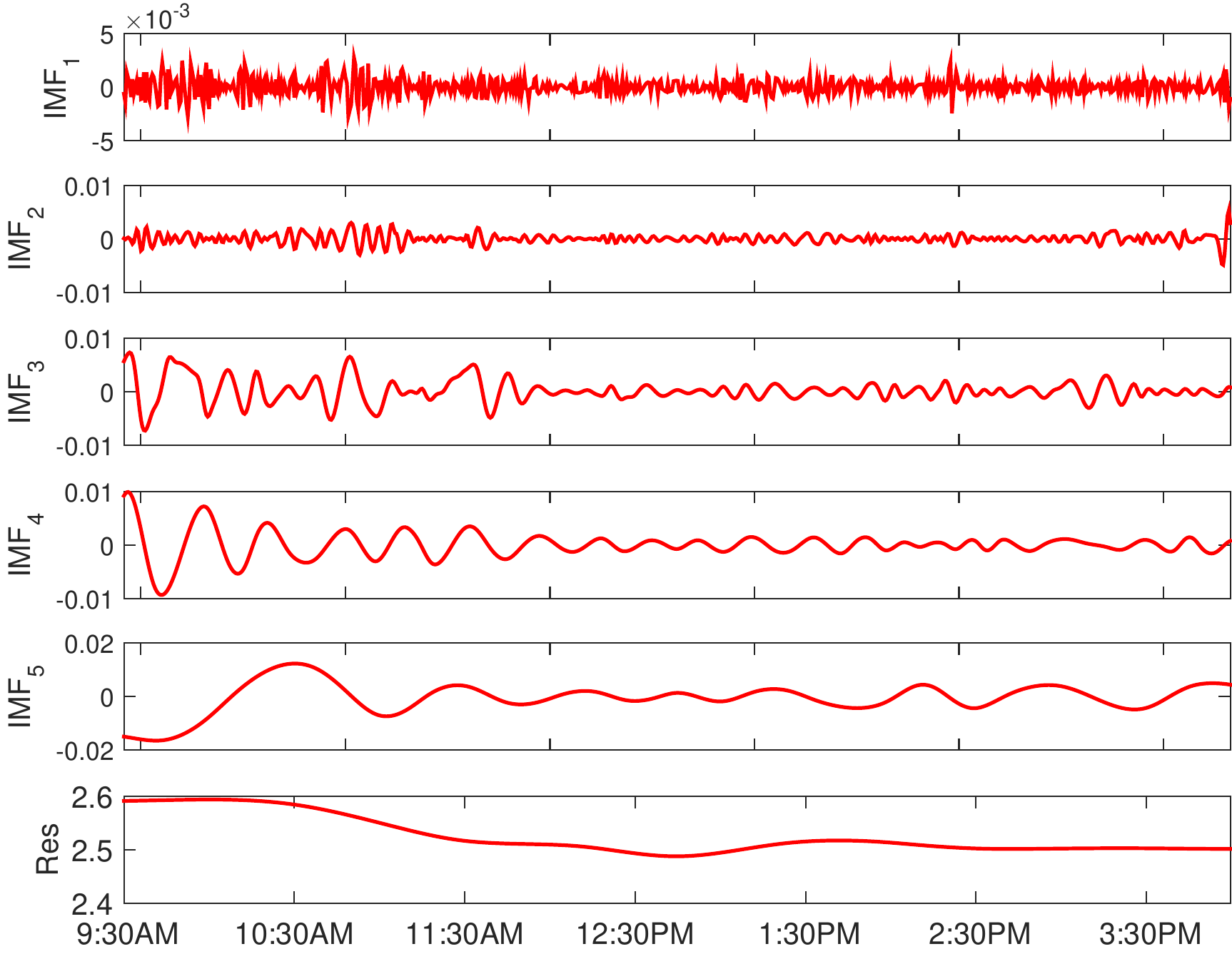}
					  \caption{VIX index.}
					  \label{fig:IMF3}
			  \end{subfigure}				  
			  \caption{IMFs of the stock market indices and the volatility index, example for July $18^{th}$ 2014. From top to bottom IMF1...IMF4 and residue. X-axes time from 9:30 to 16:00.} 
			  \label{fig:IMF3TS}
	  \end{figure}

Let us exemplify the intraday  analysis of correlation on a randomly chosen day: July $18^{th}$ 2014.  Figure  \ref{fig:3TS} displays the  logarithm of	 prices for the three indices.
Applying the  EMD to each time series, we obtained five IMFs and a residue which are reported in Figure \ref{fig:IMF3TS}.  The  oscillating period of each IMFs is calculated  by dividing the total number of points by the number of peaks, with rounded values  reported in Table \ref{tab:PeriodSingleDay}.

 \begin{table}[htbp]
   \centering
	 \begin{tabular}{ccccccc}
	 \toprule
	 \textbf{Index } & \textbf{$\IMF_1$} & \textbf{$\IMF_2$} & \textbf{$\IMF_3$} & \textbf{$\IMF_4$} & \textbf{$\IMF_5$} & \textbf{Residue} \\
	 \midrule
	 \textbf{S\&P} & 4	   & 8	   & 20	   & 44	   & 88	   &  --\\
	 \textbf{IPC} & 4	 & 8	 & 16	 & 40	 & 88	 & -- \\
	 \textbf{VIX} & 4	  & 8	  & 20	  & 40	  & 88	  & -- \\
	 \bottomrule
	 \end{tabular}%
	 \caption{Oscillating period for the IMFs  shown in Figure \ref{fig:IMF3TS} and calculated	by dividing the total number of points by the number of peaks, example for July $18^{th}$ 2014.}
   \label{tab:PeriodSingleDay}%
 \end{table}%

 \subsubsection{Time-scale-dependent correlation, example for July $18^{th}$ 2014}	 
\begin{figure}[h!]
		   \centering
		   \begin{subfigure}[t]{0.49\textwidth}
					   \centering
						\captionsetup{justification=centering}
				   \includegraphics[height=2 in]{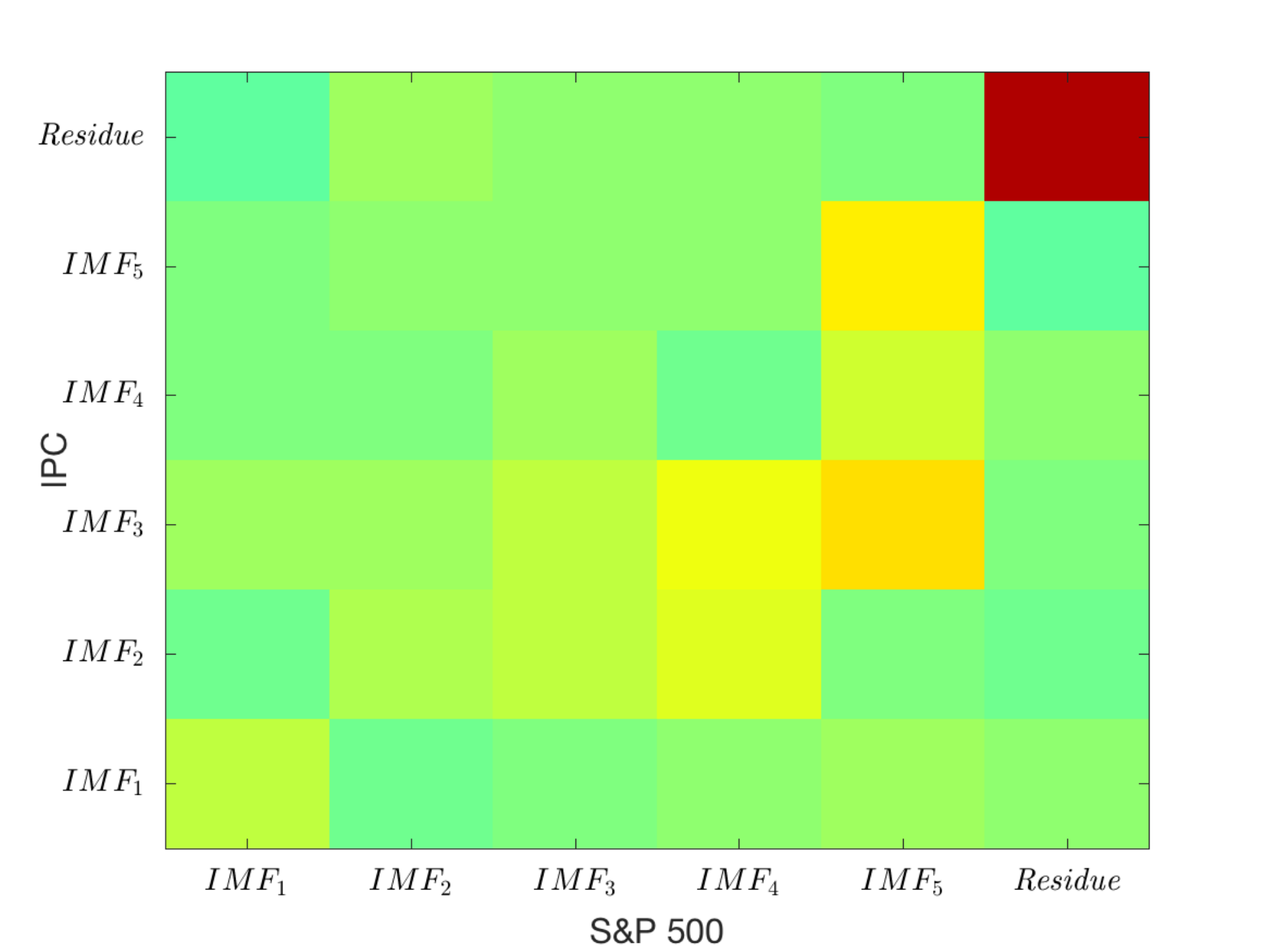}
				   \caption{ S\&P 500 index versus IPC index.}
				   \label{fig:IMF_SP_IPC}
		   \end{subfigure}%
		   \begin{subfigure}[t]{0.49\textwidth}
					   \centering
				   \includegraphics[height=2 in]{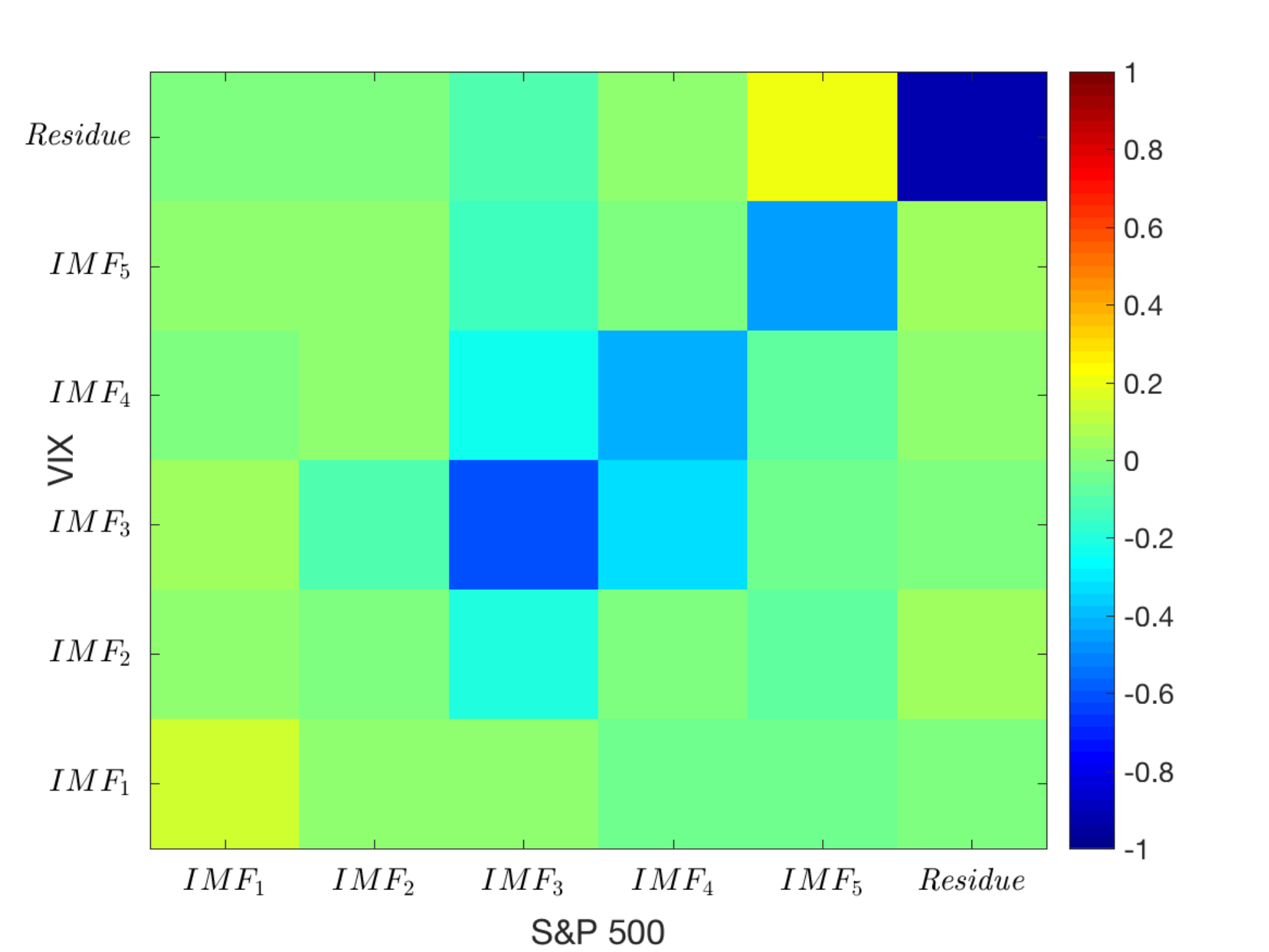}
				   \caption{S\&P 500 index versus  VIX index. }
				   \label{fig:IMF_SP_VIX}
		   \end{subfigure} 
		   \caption{Time-scale-dependent correlation structures, example for July $18^{th}$ 2014.} 
		   \label{fig:IMFCORR}
   \end{figure}

We computed the time-scale-dependent correlation  by means of Equation \eqref{eq:IMFCORR}.
The results are represented as a matrix of pairwise correlations between the IMFs where   the magnitude of the correlation is visually represented  by a color-map.  
Figure \ref{fig:IMF_SP_IPC} shows  the	correlation matrix between the	S\&P 500  and the IPC indices.
We observe positive correlations with mostly larger values on the diagonal (same time-scale components indices) with an increasing magnitude for increasing IMF time-scale.
 The  correlation between	 the S\&P and the VIX indices reveals instead negative values.  For these time series, we also observe higher correlations in the diagonal elements which are increasing with increasing time-scale (see Figure \ref{fig:IMF_SP_VIX}).

\subsubsection{Time-dependent correlation, example for July $18^{th}$ 2014}
\begin{figure}[h!]
				\centering
				\begin{subfigure}[t]{0.5\textwidth}
							\centering
							   \captionsetup{justification=centering}
						\includegraphics[width=3.3 in,	height= 3.4 in]{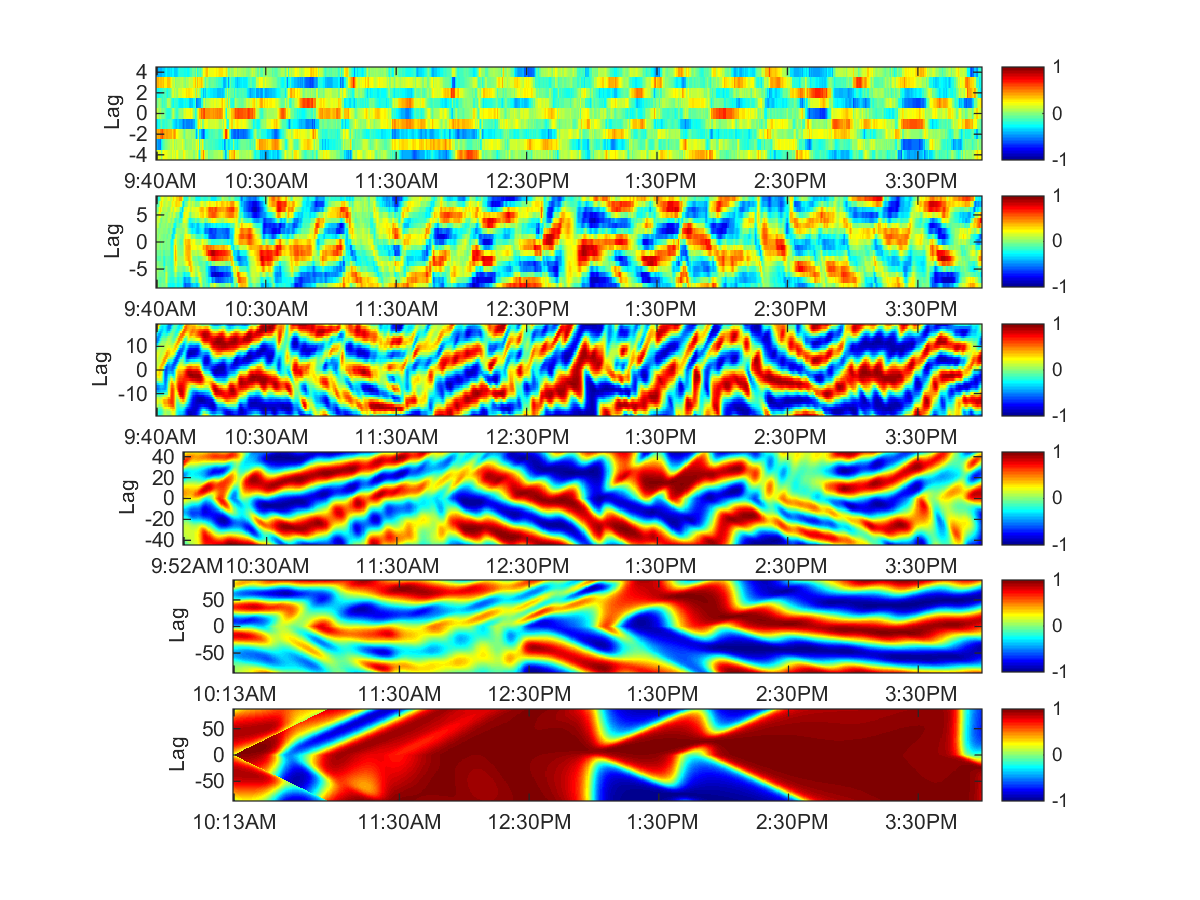}
						\caption{S\&P 500 index versus IPC index.}
						\label{fig:CROSS_SP_IPC}
				\end{subfigure}%
				\begin{subfigure}[t]{0.5\textwidth}
							\centering
						\includegraphics[width=3.3 in, height= 3.4 in]{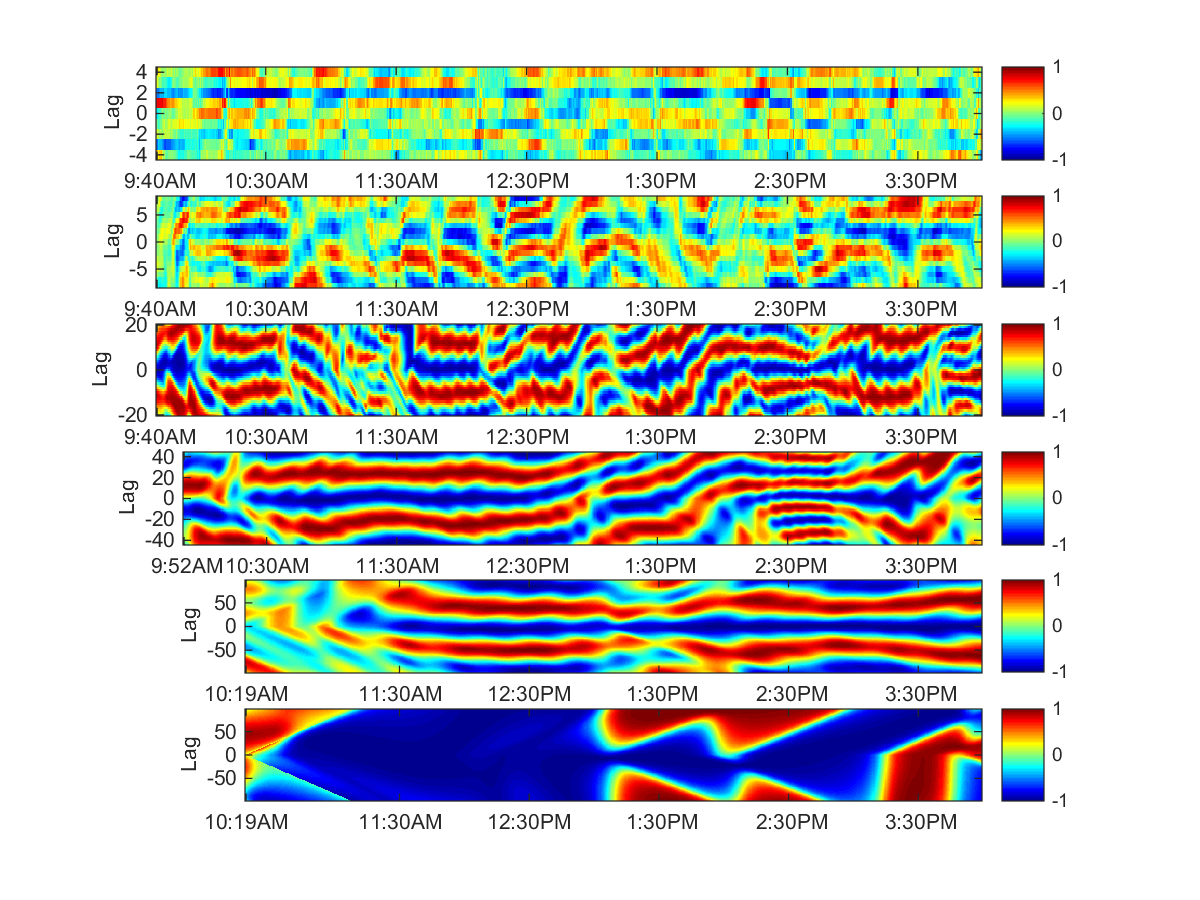}
						\caption{S\&P 500 index versus VIX index. }
						\label{fig:CROSS_SP_VIX}
				\end{subfigure} 
				\caption{Intraday time-dependent correlation, example for July $18^{th}$ 2014.}
				\label{fig:CROSSCOR}
		\end{figure}

   We also estimated the time-dependent lagged correlation	between	 the S\&P 500 and the  IPC indices by using Equation \eqref{eq:WLCross}.  
   These correlations  are  represented  in Figure \ref{fig:CROSS_SP_IPC}  as a color-map matrix in which each column represents a successive window and each row represents a specific time-lag. 
	 The intraday  correlation values are reported after  $W$ observations, with $W$ the size of the rolling-window. In this way, the size of the correlation matrix  is reduced according to the applied window. 
	   
Lags are limited to  $\lambda \le \max\left(P_{X_i},P_{Y_i}\right)$, with $P_{X_i}$ and $P_{Y_i}$ denoting the oscillating period of $\IMF_i^X$ and $\IMF_i^Y$, respectively. 
Choosing $\lambda$ larger than the oscillating period  results in repetitive patterns in the correlation structure. On the other hand, a shorter time-lag may not reveal some correlations.
The	 window size is set at $W = \max({\lambda,20})$. 

From Figure	 \ref{fig:CROSS_SP_IPC},  it is difficult to identify	correlations patterns for the  highest time-scale IMFs. However,	 for  IMFs with lower time-scale,  $\IMF_2,$ $\ldots,$ $\IMF_5$, we observe	intervals of stronger correlations  characterized by the nature of the oscillating IMFs, i.e., we observe lapses of positive correlation lagged in time by negative values of	 correlation, making the  lead-lag relation between the IMFs almost symmetric with respect to the zero lag.

Figure \ref{fig:CROSS_SP_VIX} shows the	 correlation matrices for	the S\&P 500   and the VIX indices. 
Contrary to the correlation between the S\&P 500 and the IPC indices, the correlation between the S\&P and the VIX indices is negative at all frequencies and during the entire trading day.
At the highest time-scale, $\IMF_1$, we observe a clear pattern of negative correlation at  lag $\lambda=2$ (1 min), indicating that  the S\&P 500 leads the VIX index by 1 minute. 
When correlating the residue components, we observe a dominant blue band, indicating a negative correlation region  (a similar red band is observed for the correlation between the  S\&P 500 and the IPC indices). Such a band could be attributed to the linear and non-stationary characteristics of the residues.

\subsection{Intraday correlation,  analysis on the complete data set}
Proceeding in the same way as in the previous example on July $18^{th}$ 2014, we decomposed each daily time series into five IMFs and a residue. 
We computed the time-scale-dependent correlation and the time-dependent correlation for each of the 184 days available in the data set.

\subsubsection{Time-scale-dependent	correlation}
\begin{figure}[ht]
	\centering	
\includegraphics[width=4.6 in, height= 2.9 in]{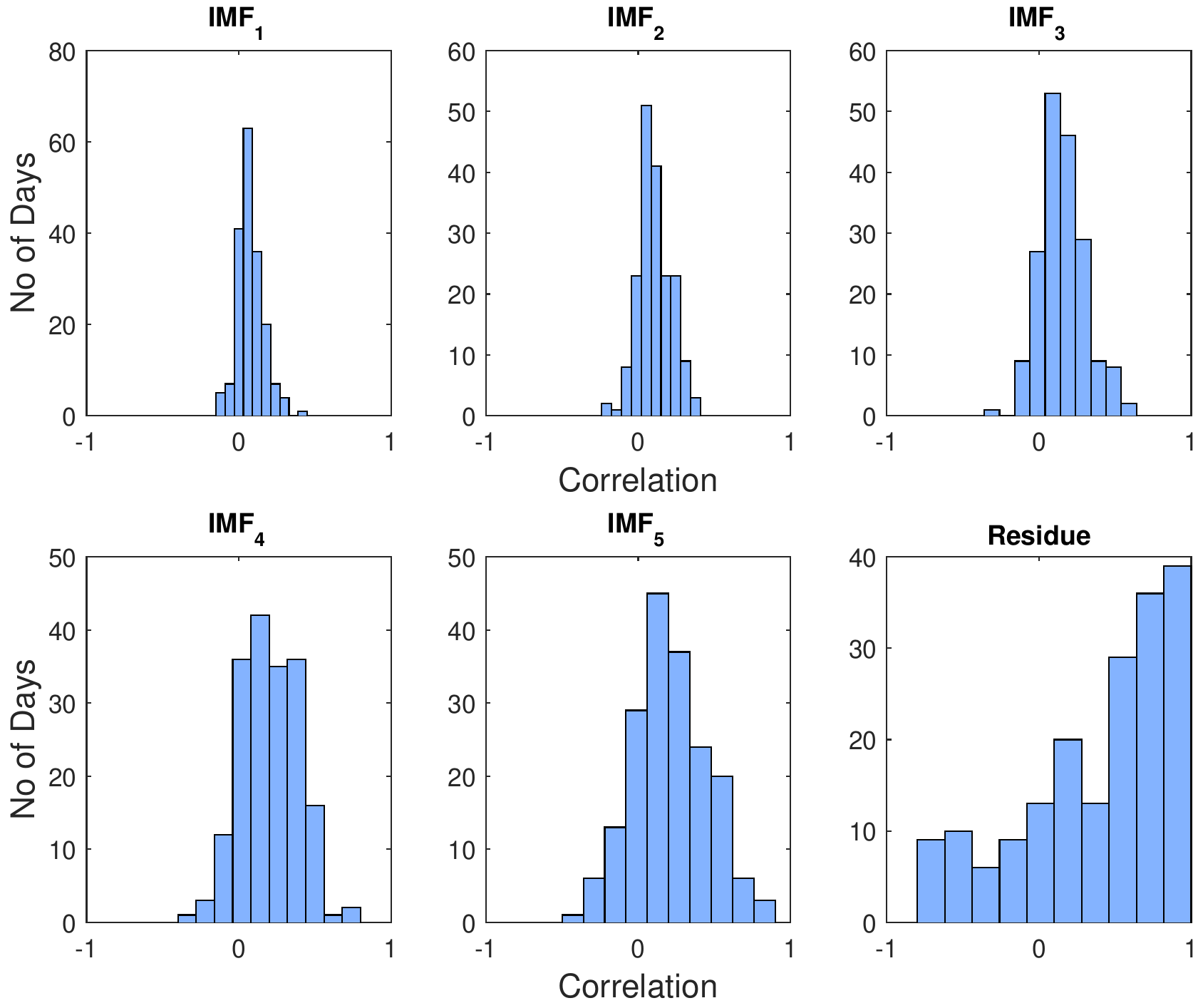}
			\caption{Distribution of the time-scale-dependent correlation  between the IMFs of the S\&P	500 index and the IMFs of the  IPC index.}
			\label{fig:HistSP_IPC}
 \end{figure}

\begin{figure}[ht]
	\centering	
\includegraphics[width=4.6 in, height= 2.9 in]{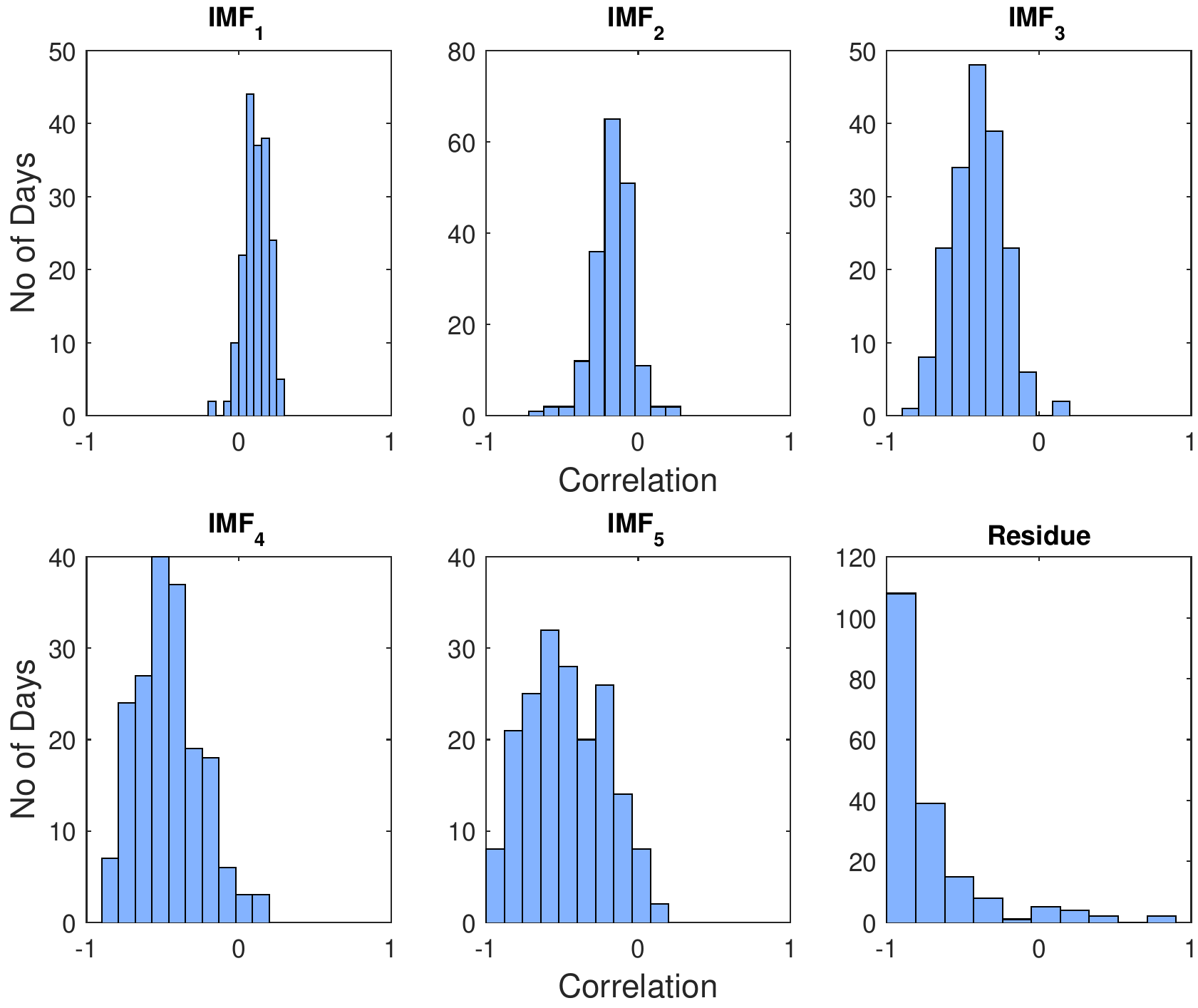}
			\caption{Distribution of the time-scale-dependent correlation  between the IMFs of the S\&P	500 index and the IMFs of the  VIX index.}
			\label{fig:HistSP_VIX}
 \end{figure}

The statistics for the time-scale dependent correlation between the IMFs with the same time-scale index for all trading days are reported in  histograms.  In  Figure \ref{fig:HistSP_IPC}, we report histograms	for the S\&P 500  and the IPC indices and in     Figure  \ref{fig:HistSP_VIX}  for the  S\&P and the VIX indices. 
We observe prevalently positive correlations for the S\&P 500  and the IPC components and instead prevalently negative correlations for the S\&P 500  and the VIX.
The histograms reveal significant deviations from zero for all the components with larger positive or negative correlations for components with longer  time-scales.


\begin{figure}[h!]
		   \centering
		   \begin{subfigure}[t]{0.48\textwidth}
					   \centering
						  \captionsetup{justification=centering}
				   \includegraphics[width=2.5 in]{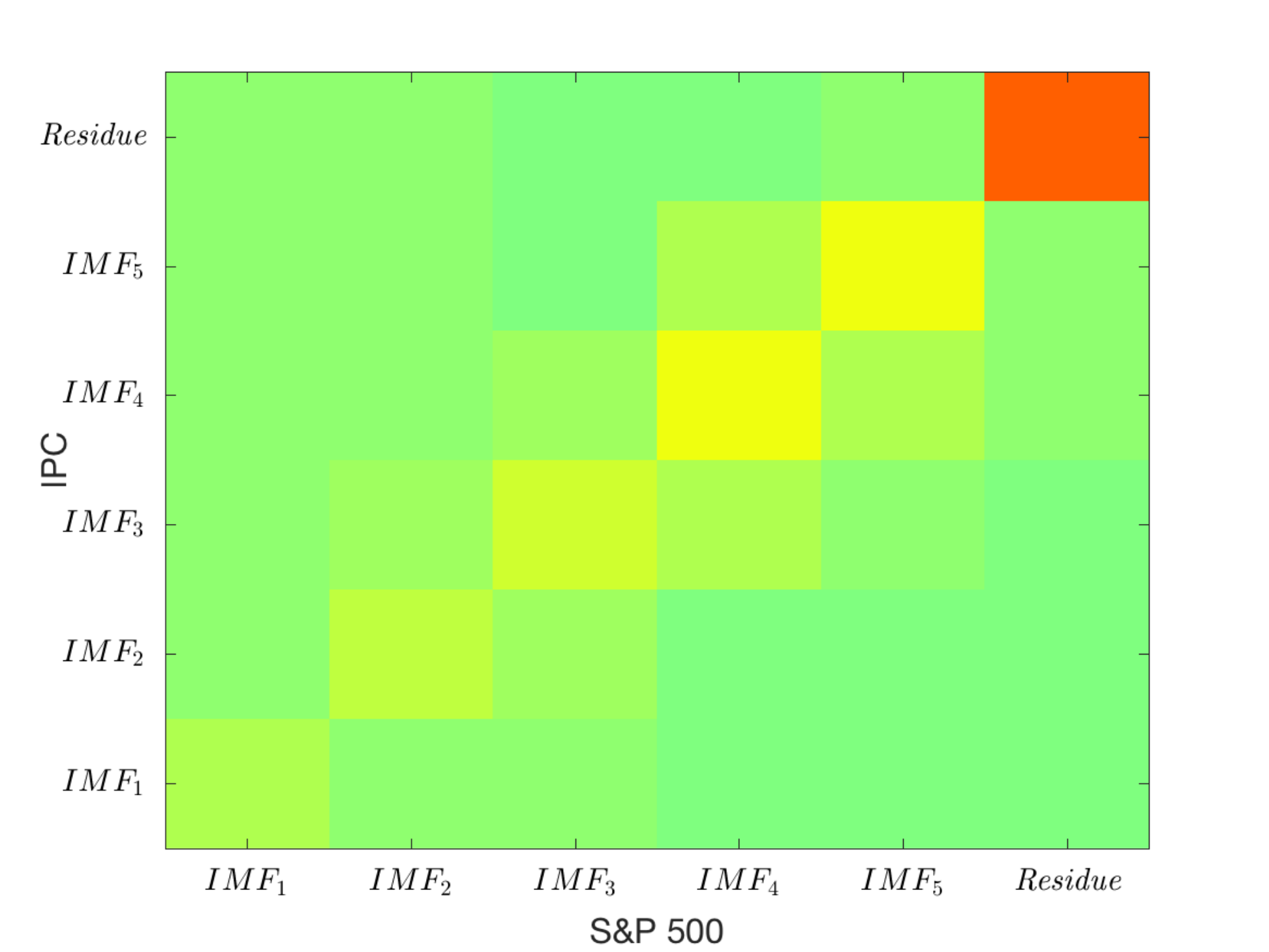}\\
				   \includegraphics[width=2.5 in]{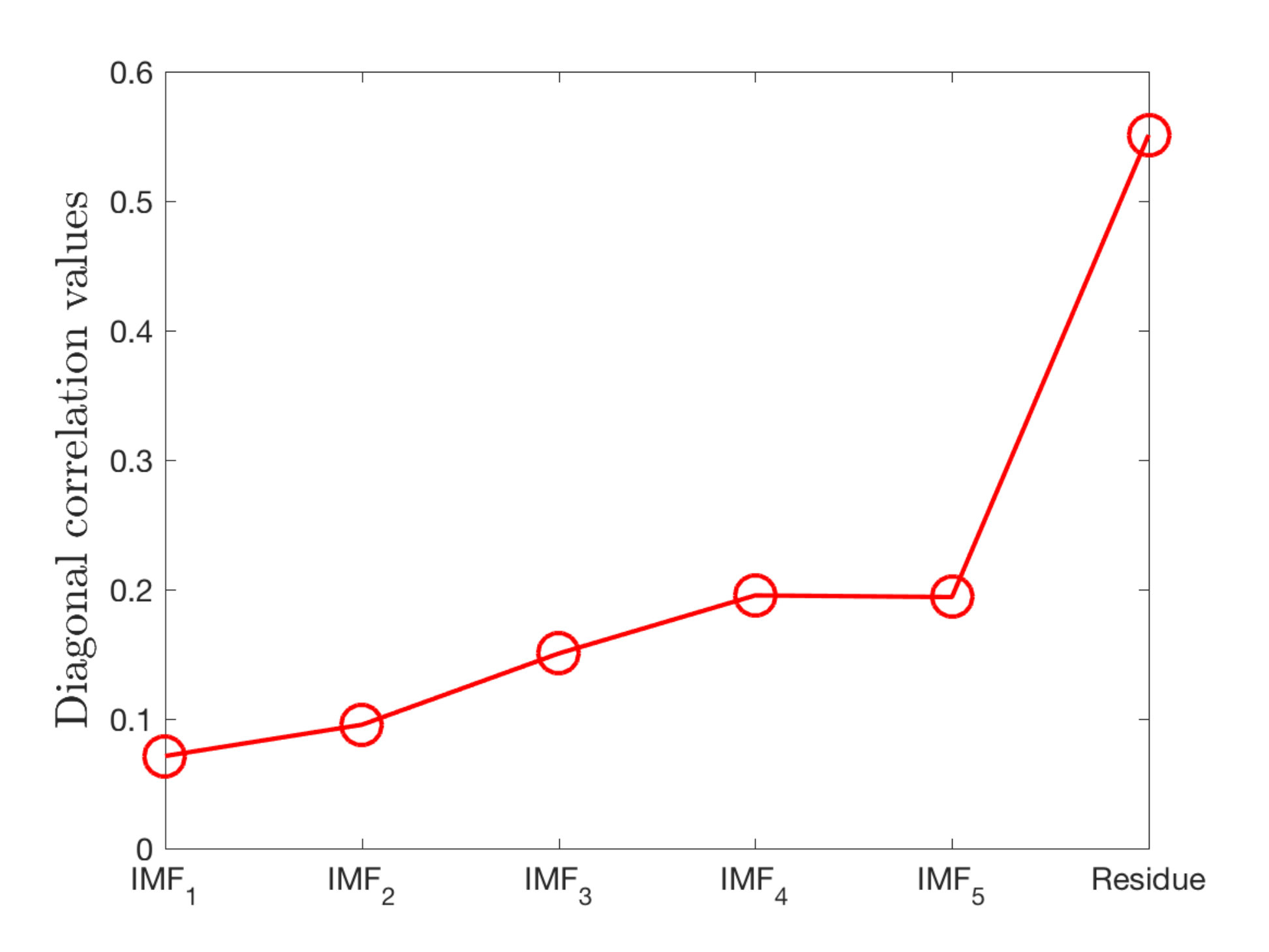} 
				   \caption{S\&P 500 index and the IPC index.}
				   \label{fig:medianIMF_SP_IPC}
		   \end{subfigure}%
		   \begin{subfigure}[t]{0.48\textwidth}
					   \centering
				   \includegraphics[width=2.5 in]{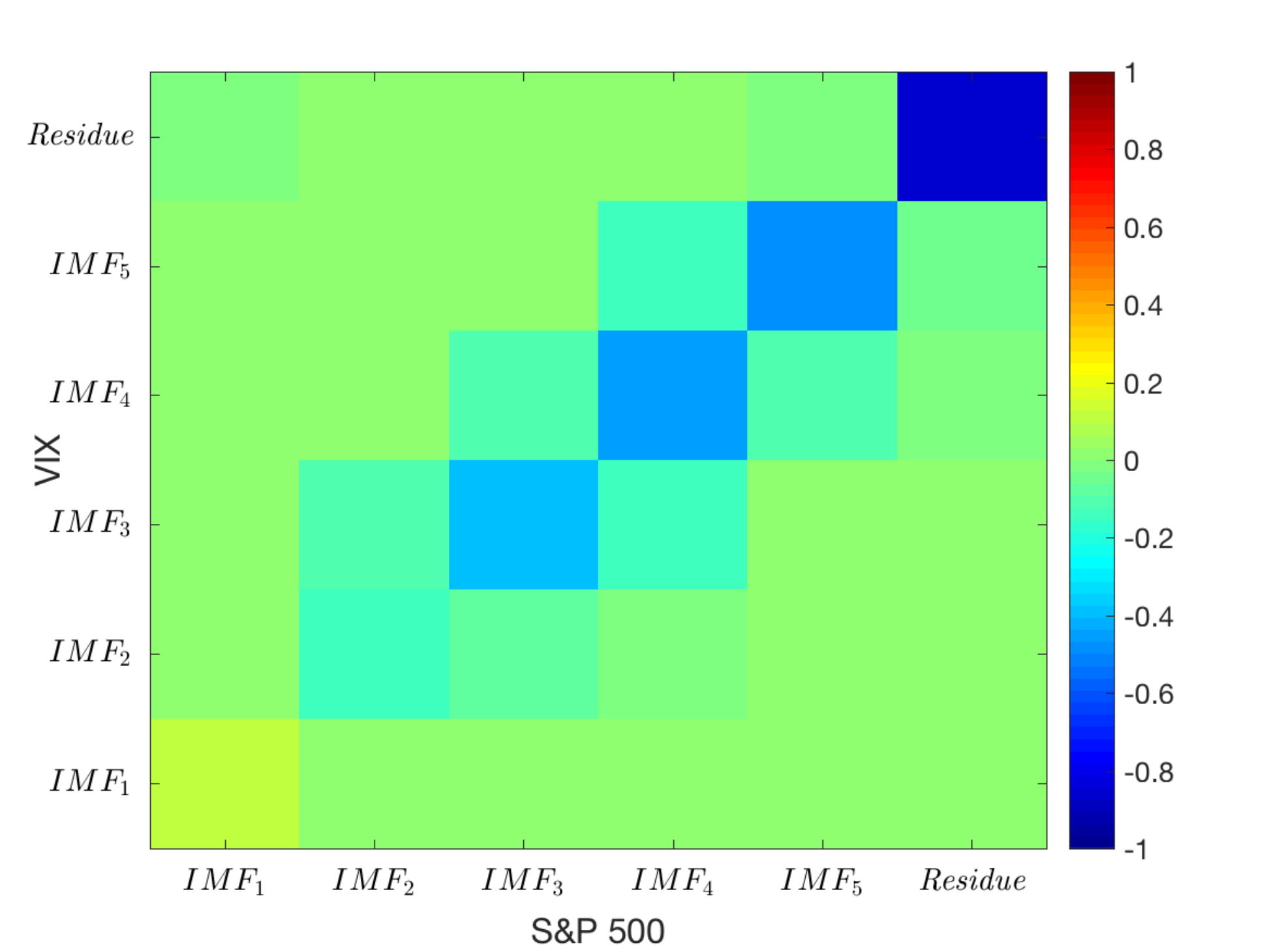}\\
				   \includegraphics[width=2.5 in]{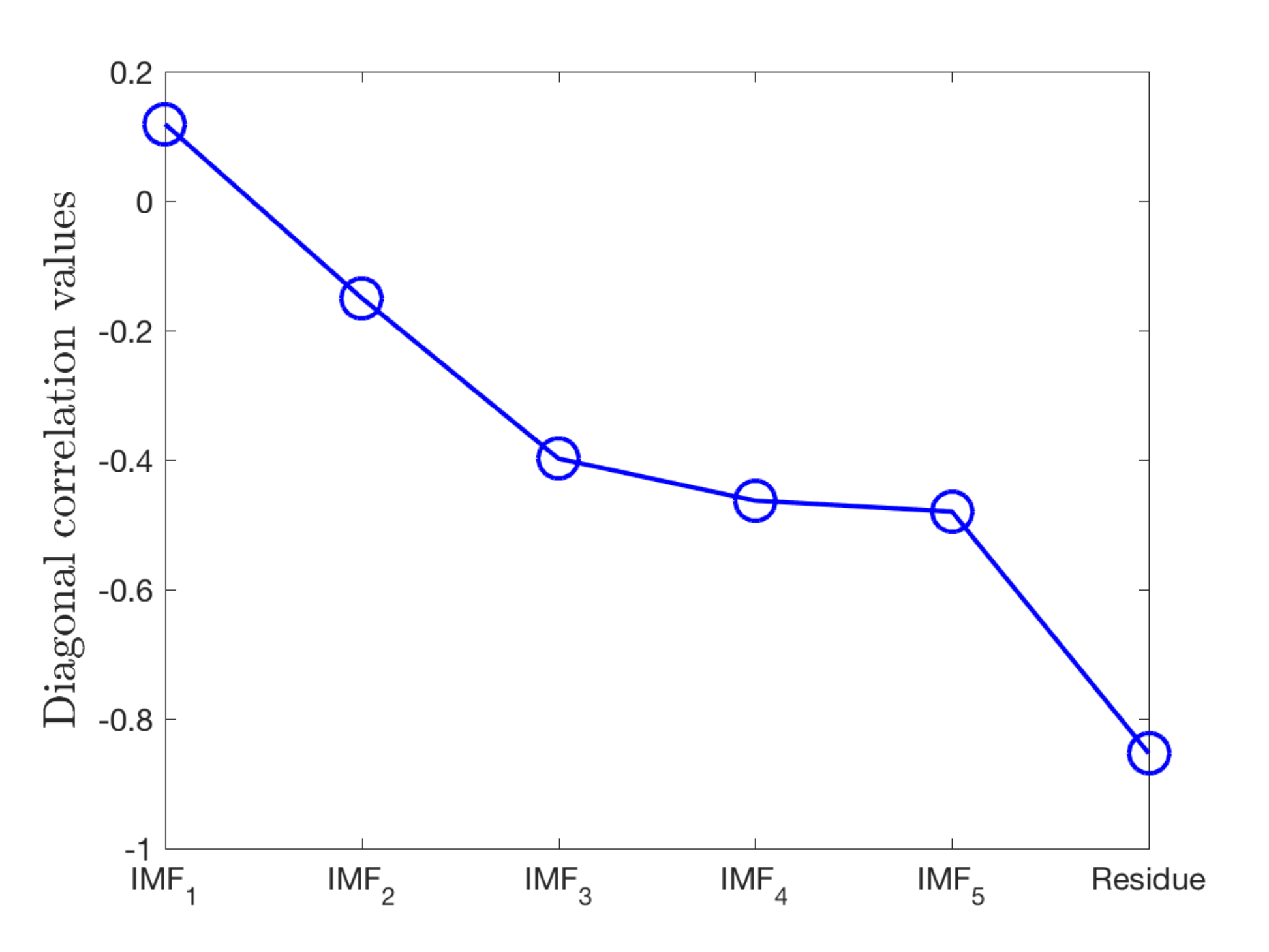}\\
				   \caption{S\&P 500 index and the VIX index. }
				   \label{fig:medianIMF_SP_VIX}
		   \end{subfigure} 
		   \caption{
		   Sample median of the time-scale-dependent correlation matrices over the time period from September 2013 to July 2014.
		   The color-map matrices above are the cross correlations between components at  different time-scales; the plots below report the values of the elements in the diagonal with same time-scales indices.} 
		   \label{fig:medianIMF}
   \end{figure}

Correlations between the IMF components with different time-scales indices are reported in Figure \ref{fig:medianIMF} where the sample median correlations are also reported. 
We use the sample	median of the distribution since this statistic is not influenced by outliers. 
The case S\&P and IPC is shown in Figure \ref{fig:medianIMF_SP_IPC} and the S\&P and  VIX in  Figure \ref{fig:medianIMF_SP_VIX}. 
The color-map matrices at the top are the median correlations  between different time-scales whereas the plots below are the values of the diagonal elements (components with same-time scale indices).

\subsubsection{Rolling window analysis and lag relations}	 
We	analysed  the median of the time-dependent correlation matrices, computed 
as reported in Equation \ref{eq:WLCross}, over rolling windows and with lags. 
The window sizes and the time lags which were used are reported in Table \ref{tab:AveragePeriod}.

\begin{table}[htbp]
 \centering
\resizebox{8 cm}{!}{
   \begin{tabular}{ccccc}
   \toprule
   \textbf{Component} & \multicolumn{2}{c}{\textbf{S\&P vs IPC}} & \multicolumn{2}{c}{\textbf{S\&P vs VIX}} \\
   \midrule
   \textbf{} & \textbf{Lag} & \textbf{Window} & \textbf{Lag} & \textbf{Window} \\
   \textbf{$\IMF_1$} & 4	 & 20	 & 4	 & 20 \\
   \textbf{$\IMF_2$} & 9	 & 20	 & 9	 & 20 \\
   \textbf{$\IMF_3$} & 19	 & 20	 & 21	 & 21 \\
   \textbf{$\IMF_4$} & 44	 & 44	 & 48	 & 48 \\
   \textbf{$\IMF_5$} & 110	 & 110	 & 124	 & 124 \\
   \textbf{Residue} & 110	& 110	& 124	& 124 \\
   \bottomrule
   \end{tabular}%
   }
	 \caption{Average of the number of lags and	 the size of the rolling-window used for the time-dependent correlation analysis.}
 \label{tab:AveragePeriod}%
\end{table}%

\begin{figure}[h!]
				\centering
				\begin{subfigure}[t]{0.48\textwidth}
							\centering
						\includegraphics[width=3.2 in, height= 3.4 in]{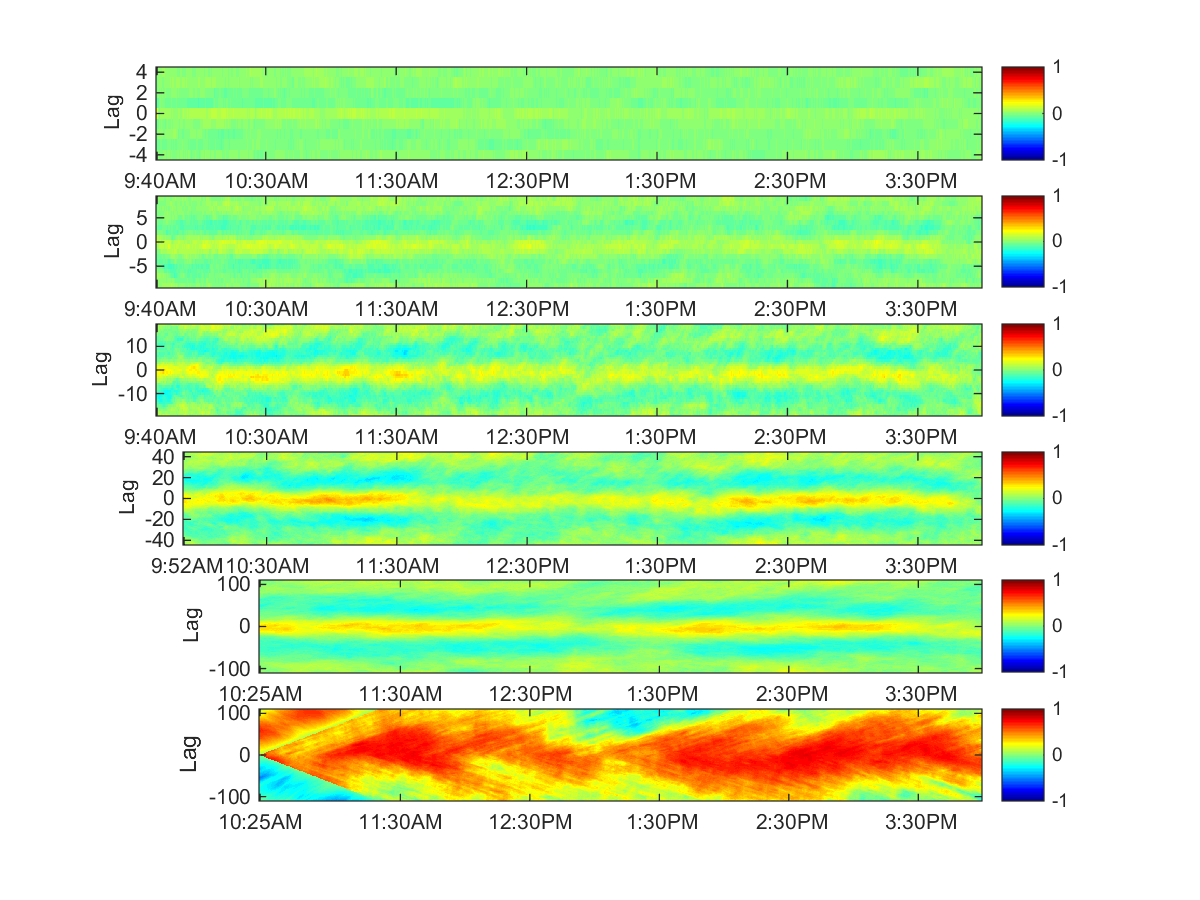}
						\caption{S\&P 500 index versus IPC index.}
						\label{fig:medianCROSS_SP_IPC}
				\end{subfigure}%
				\begin{subfigure}[t]{0.48\textwidth}
							\centering
							\captionsetup{justification=centering}
						\includegraphics[width=3.2 in, height= 3.4 in]{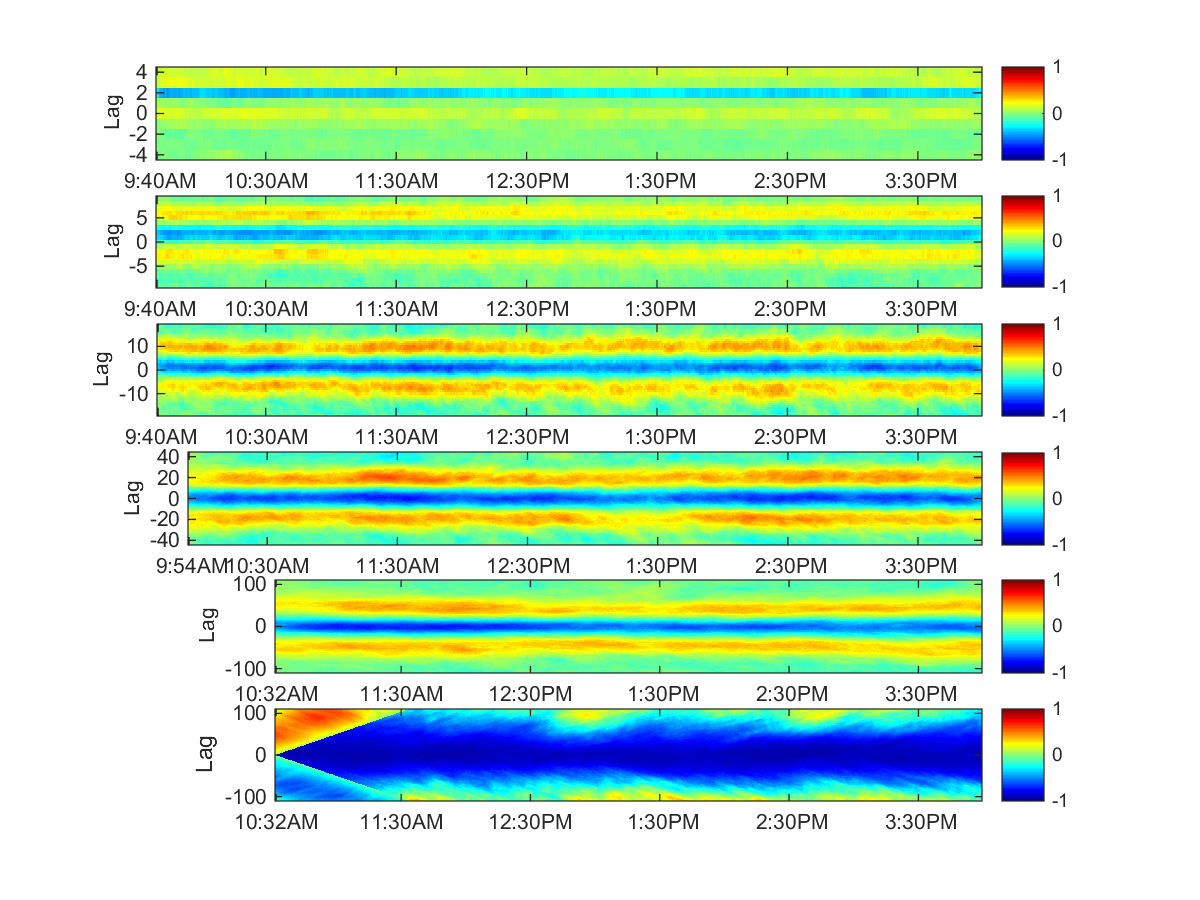}
						\caption{S\&P 500 index versus	VIX index.}
						\label{fig:medianCROSS_SP_VIX}
				\end{subfigure} 
				\caption{Sample median of the time-dependent correlation matrices over the time period from September 2013 to July 2014.}
				\label{fig:medianCROSS}
		\end{figure}

The	 median, time-varying, lagged correlation matrix (Eq.\ref{eq:WLCross}) between the S\&P 500 and the IPC indices is displayed in Figure \ref{fig:medianCROSS_SP_IPC}.
Overall, we observe relatively small correlations with little  lead-lag relationships at all time-scales with larger values for the last two components and the residual (bottom panels).
We observe patterns in the intraday activity with less persistent correlations around the middle of the day.
  
More intense negative correlation is observed between the S\&P 500 and the VIX indices, reported in Figure \ref{fig:medianCROSS_SP_VIX}.	
Interestingly, in this case, we observe significant lagged correlations at small time scales ($\IMF_1$, $\IMF_2 $ and $\IMF_3$) with the S\&P 500 leading the VIX at one minute lag ($\lambda=2$) with a stable pattern across the day.
This indicates that consistently changes in the S\&P 500 are followed  by changes in the VIX after about  1 min and in the opposite direction (negative correlations).

\section{Discussions and conclusions}
\label{sec:CorrConclusion}	

In this paper we propose a simple approach which shows that Empirical Mode Decomposition can be used to investigate the correlation between time series at different time-scales.
This expands the concept of correlations to a higher-dimensional level.
We observed that, although most of the correlation is between components of the same time-scales, there are  some significant correlations also between components of different time-scales. 
A dynamical analysis performed over rolling windows of 30 seconds    shows that correlations' patterns are both time and time-scale dependent. 
We uncovered lead-lag relations within components with the discovery of a persistent and significant 1 min negative coupling between the S\&P 500 and the VIX indices which can have practical relevance for trading strategies and risk modeling.

The methodology we introduced in this paper and our findings are consistent with the coherence measure obtained with the wavelet transform~\cite{Vacha2012241}. 
However, given the simplicity of the EMD method and its adaptability to different time series without needing to specify  any filter function, we believe that the proposed correlation measures offer a simpler, computationally more efficient and easier to interpret approach. 

The measures proposed in this paper and in particular Eqs. \ref{eq:IMFCORR}, \ref{eq:WLCross} are the simplest generalisations of the linear correlation measure to include time-scale components.  
We chose them as a natural extension of the cross-correlation concept. 
However, there are  some aspects of  the present approach that would be interesting to further investigate in future works.
For instance, Eqs. \ref{eq:IMFCORR}, \ref{eq:WLCross} perform averages over the variables, but time-series with different oscillation scales lead to different averages even if the (scaled) nature of the variable is the same. 
This is probably penalising the values associated with high frequency components that we indeed  observe to be consistently smaller.

 \section*{Acknowledgement}
  The authors wish to thank Bloomberg for providing the data. NN would like to acknowledge the financial support from Conacyt-Mexico. TDM wishes to thank the COST Action TD1210 for partially supporting this work. TA \& TDM wish to thank the Systemic Risk Centre at LSE.

\newpage

\bibliographystyle{unsrt}

\end{document}